\documentclass[showpacs,preprintnumbers,amsmath,amssymb,12pt,floatfix,epsfig]{revtex4}
\usepackage{dcolumn}
\usepackage{bm}
\usepackage[dvips]{color}
\usepackage{graphicx}

\usepackage{graphicx}
\usepackage{epsfig}

\baselineskip=24ptplus.5ptminus.2pt \vspace*{0.1 in} \large

\def\prl#1{Phys.\ Rev.\ Lett.\ {\bf #1}}
\def\pr#1{Phys.\ Rev.\ {\bf #1}}

\def\pl#1{Phys.\ Lett.\ {\bf #1}}

\def\mo100{${}_{42}^{100}Mo_{58}$}

\def\0nu{0$\nu$}

\def\be{\begin{equation}}
\def\ee{\end{equation}}
\def\Be{\begin{eqnarray}}
\def\Ee{\end{eqnarray}}
\def\ba{\begin{array}}
\def\ea{\end{array}}

\begin{document}
\title{\bf Neutrino reactions on $^{138}$La and $^{180}$Ta via charged and neutral currents
by the Quasi-particle Random Phase Approximation (QRPA)}
\author{
Myung-Ki Cheoun${^{1)}}$\footnote{Corresponding author :
cheoun@ssu.ac.kr}, Eunja Ha$^{1)}$, T. Hayakawa$^{2)}$, Toshitaka
Kajino$^{3,4)}$, Satoshi Chiba$^{5)}$ }
\address{
1) Department of Physics, Soongsil University, Seoul 156-743,
Korea
\\
2) Kansai Photon Science Institute, Japan Atomic Energy Agency,
Kizu, Kyoto 619-0215, Japan  \\
3) National Astronomical Observatory, Mitaka, Tokyo 181-8589,
Japan \\
4) Department of Astronomy, Graduate School of Science, University
of Tokyo, 7-3-1 Hongo, Tokyo 113-0033, Japan \\
5) Advanced Science Research Center, Japan Atomic Energy Agency,
2-4 Shirakata-shirane, Tokai, Ibaraki 319-1195, Japan }

\def\ref#1{$^{#1)}$}

\begin{abstract}
Cosmological origins of the two heaviest odd-odd nuclei,
$^{138}$La and $^{180}$Ta, are believed to be closely related to
the neutrino-process. We investigate in detail neutrino-induced
reactions on the nuclei. Charged current (CC) reactions,
$^{138}$Ba$ ( \nu_e , e^{-}) ^{138}$La and $^{180}$Hf$ ( \nu_e ,
e^{-}) ^{180}$Ta, are calculated by the standard Quasi-particle
Random Phase Approximation (QRPA) with neutron-proton pairing as
well as neutron-neutron, proton-proton pairing correlations. For
neutral current (NC) reactions, $^{139}$La$ ( \nu , \nu^{'} )
^{139}${La}$^*$ and $^{181}$Ta$ ( \nu , \nu^{'}) ^{181}$Ta$^*$, we
generate ground and excited states of odd-even target nuclei,
$^{139}$La and $^{181}$Ta, by operating one quasi-particle to
even-even nuclei, $^{138}$Ba and $^{180}$Hf, which are assumed as
the BCS ground state. Numerical results for CC reactions are shown
to be consistent with recent semi-empirical data deduced from the
Gamow-Teller strength distributions measured in the ($^{3}$He, t)
reaction. Results for NC reactions are estimated to be smaller by
a factor about 4 $\sim$ 5 rather than those by CC reactions.
Finally, cross sections weighted by the incident neutrino flux in
the core collapsing supernova are presented for further
applications to the network calculations for relevant nuclear
abundances.
\end{abstract}



\maketitle

\section{Introduction}
Astrophysical origins of the two heaviest odd-odd nuclei,
${}_{57}^{138}$La and ${}_{73}^{180}$Ta, have been discussed over
the last 30 years \cite{Beer81,Yokoi83,Woosley78,Woosley90}.
Destruction rates of the odd-odd nuclei by particle- or
photon-induced reactions are generally larger than production
rates. These two isotopes are shielded against both ${\beta}^{-}$
and ${\beta}^{+}$ decays by stable isobars. Thus nucleosynthesis
calculations by the slow ($s$), rapid ($r$) and $\gamma$ processes
usually underproduced solar abundances of these nuclei.

In a core collapsing supernova, incident neutrino ($\nu$)
(antineutrino (${\bar \nu}$)) energies and flux emitted from a
proto-neutron star \cite{Woosley90,yoshida08} are presumed to be
peaked from a few to tens of MeV energy region by considering the
Fermi-Dirac distribution characterized by the temperature in each
astrophysical site \cite{Kolbe03-a}. The $\nu ({\bar \nu})$
naturally interacts with the nuclei inside the dense matter and
usually proceeds via two-step processes, {\it i.e.} target nuclei
are excited by incident $\nu ({\bar \nu}) $ and decay to lower
energy states with the emission of some particles
\cite{Kolbe03-a}. The excitation occurs through various
transitions, {\it i.e.} super allowed Fermi ($J^{\pi} = 0^+)$,
allowed Gamow-Teller (GT) ($J^{\pi} = 1^+)$, spin dipole ($J^{\pi}
= 0^- , 1^-, 2^-)$ and other higher multipole transitions
\cite{Suzuki06,Ring08}. Consequently, the $\nu$-process,
neutrino-induced reactions on related nuclei in core collapse
supernovae \cite{Woosley90,Heg,yoshida08,Suzuki09,Wana06}, are
treated as important input data for the network calculations to
explain astrophysical origins and abundances of relevant nuclei
\cite{Suzuki06}.

Before presenting our results, we briefly summarize recent
theoretical and experimental status about the abundances of the
two heaviest odd-odd nuclei by following Refs.
\cite{Heg,RCNP,Haya}. Since $^{138}$La lies on the
neutron-deficient region, it is bypassed by successive neutron
captures, $s$- and $r$-processes \cite{Haya}. In the 15 solar mass
supernova (SN) model \cite{Heg}, the $\gamma$-process by
photo-disintegration and the $\nu$-process by neutral current
(NC), $^{139}$La$ ( \nu , \nu^{'} ) ^{139}${La}$^*$, followed by a
neutron emission, are shown to account for 32 \% of the
${}^{138}$La abundance \cite{Anders89}. Remainder comes from the
charged current (CC) reaction, $^{138}$Ba$ ( \nu_e , e^{-})
^{138}$La. By adding the CC reaction to the SN model at $T_{\nu}$=
4 MeV, one can reproduce the abundance within about 10 \%
uncertainty. This reproduction could be a reasonable result if we
consider the ambiguities inherent in the SN model.

For $^{180}$Ta, the $\gamma$-process and the $\nu$-process via NC,
$^{181}$Ta$ ( \nu , \nu^{'}) ^{181}$Ta$^*$, already overproduce
the abundance about 15 \% \cite{Anders89}. Addition of $^{180}$Hf
$ ( \nu_e , e^{-}) ^{180}$Ta reaction at $T_{\nu}$= 4 MeV
overestimates the abundance about 3 times \cite{Heg}. The
overestimation is thought to be originated from the unique feature
that $^{180}$Ta has a meta-stable isomer $9^-$ state whose
half-life is larger than $10^{15}$ yr, while the ground $1^+$
state is beta unstable with a half life of only 8.15 hr
\cite{Haya}. Linkage transitions between the states could play an
important role in reducing the overproduction \cite{Haya10}.

The SN model simulation for the abundances exploited the
$\nu$-induced reaction data calculated by the Random Phase
Approximation (RPA) model \cite{Heg}. To reduce the ambiguities
persisting on nuclear models, semi-empirical deduction of relevant
CC reactions is performed recently by using the GT strength
obtained from $^{138}$Ba ($^{3}$He , $t$)$^{138}$La and $^{180}$Hf
($^{3}$He , $t$)$^{180}$Ta reaction data \cite{RCNP}. The
$^{138}$La abundance by the data turns out to be about 15 \%
higher than that by the theoretical data in the 15 solar mass SN
model at $T_{\nu}$= 4 MeV \cite{Heg}. Consequently, 25 \% of the
population is overproduced by the SN model. The overproduction of
$^{180}$Ta becomes larger about 10 \% than the SN model by the
theoretical data \cite{Heg}, {\it i.e.} the 3 times overproduction
of $^{180}$Ta is still remained to be solved.

However, for more precise estimation of nuclear abundances, one
needs more systematic calculations of neutrino reactions by
pinning down the ambiguities on nuclear models. For instance, the
uncertainty between theoretical and semi-empirical results for CC
reactions is within 10 $\sim$ 20 \% at $T_{\nu}$= 4 MeV. But the
discrepancy becomes about 2 times at $T_{\nu}$= 8 MeV, even if we
take it for granted the recipe deducing the CC reaction data from
the experimental GT strength \cite{RCNP}. Roles of NC reactions
also deserve to be detailed because they do not have any other
theoretical and empirical results to be compared with. Moreover
the NC reaction is subject to the proper description of ground
states of odd-even nuclei.

Here we report more advanced results based on the QRPA
calculation, which described very well the available neutrino
reaction data on light nuclei \cite{Ch09-1,Ch09-2}. Our results
are shown to reproduce the recent GT strength data on $^{138}$La
and $^{180}$Ta, and their semi-empirical $\nu-$induced reaction
data \cite{RCNP}. Results for NC reactions are compared with the
calculations used at the SN model \cite{Heg,RCNP}.

In Sec. II, we summarize our theoretical models for the neutrino
reactions on even-even and odd-even nuclei. Numerical results and
detailed discussion are presented at Sec. III. Finally summary and
conclusion are given at Sec. IV.

\section{Theoretical Framework}

Since our QRPA formalism for the the $\nu ({\bar \nu})$-nucleus
($\nu ({\bar \nu}) - A)$ reaction is detailed at our previous
papers \cite{Ch09-1,Ch09-2}, here we summarize two important
characteristics compared to other QRPA approaches. First, we
include neutron-proton (np) pairing as well as neutron-neutron
(nn) and proton-proton (pp) pairing correlations. Consequently,
both reactions via CC and NC are described within a framework. In
medium or medium-heavy nuclei, the np pairing is usually expected
to contribute to some extent for relevant transitions because of
small energy gaps between proton and neutron energy spaces
\cite{Ch93}.

Second, the Brueckner G matrix is employed for two-body
interactions inside nuclei by solving the Bethe-Salpeter equation
based on the Bonn CD potential for nucleon-nucleon interactions in
free space. It may enable us to reduce some ambiguities from
nucleon-nucleon interactions inside nuclei. Results by our QRPA
have successfully described relevant $\nu -$induced reaction data
for $^{12}$C, $^{56}$Fe and $^{56}$Ni \cite{Ch09-1,Ch09-2} as well
as $\beta$, 2$\nu \beta \beta$ and $0 \nu 2 \beta$ decays
\cite{Ch93}. In specific, the double beta decay is well known to
be sensitive on the nuclear structure and has more data than the
$\nu$-induced reaction data. Therefore, it could be a useful test
of nuclear models adopted for the $\nu$-process.

In our QRPA, the ground state of a target nucleus, which is
assumed as an even-even nucleus, is described by the BCS vacuum
for the quasi-particle which comprises nn, pp and np pairing
correlations. Excited states, $\vert m; J^{\pi} M \rangle$, in a
compound nucleus are generated by operating the following one
phonon operator to the initial BCS state
\begin{equation} Q^{+,m}_{JM}  = {\mathop\Sigma_{k l \mu^{'}
\nu^{'} }} [ X^{m}_{(k \mu^{'} l \nu^{'} J)} C^{+}(k \mu^{'} l
\nu^{'} J M)
 - Y^{m}_{(k \mu^{'} l \nu^{'} J)} {\tilde C}(k \mu^{'} l \nu^{'}
  J M)] ~,\end{equation}
where pair creation and annihilation operators, $C^+$ and ${\tilde
C}$, are defined as
\begin{equation}
 C^{+} (k \mu^{'} l \nu^{'} J M)  =  {\mathop\Sigma_{m_{k} m_{l}}}
C^{JM}_{j_{k} m_{k} j_{l} m_{l}} a^{+}_{l \nu^{'}} a^{+}_{k
\mu^{'}}~,~ {\tilde C}(k \mu^{'} l \nu^{'} J M)  =  (-)^{J-M} C(k
\mu^{'} l \nu^{'} J - M)~
\end{equation}
with a quasi-particle creation operator $a^{+}_{l \nu^{'}}$ and
Clebsh-Gordan coefficient $C^{JM}_{j_{k} m_{k} j_{l} m_{l}}$. Here
Roman letters indicate single particle states, and Greek letters
with a prime mean quasi-particle types 1 or 2.

If the neutron-proton pairing is neglected, quasi-particles become
quasi-proton and quasi-neutron, and the phonon operator is easily
decoupled to two different phonon operators. One is for the charge
changing reaction such as the nuclear $\beta$ decay and CC
neutrino reactions. The other is for the charge conserving
reaction such as electro-magnetic and NC neutrino reactions. The
amplitudes $X_{a {\alpha}^{'}, b {\beta}^{'}}$ and $Y_{a
{\alpha}^{'}, b {\beta}^{'}}$, which stand for forward and
backward going amplitudes from ground states to excited states,
are obtained from the QRPA equation, whose detailed derivation was
shown at Refs. \cite{Ch93,Ch09-2}

Under the second quantization, matrix elements of any transition
operator ${\cal {\hat O}}$ between a ground state and an excited
state $ | \omega ; J M >$ can be factored as follows
\begin{equation} < QRPA || {\cal {\hat O}}_{\lambda } || ~ \omega ; JM
>  =  {[\lambda]}^{-1} {\mathop{\Sigma}_{ab}} < a ||  {\cal {\hat O}}_{\lambda} || b>
<  QRPA || {[c_a^+ {\tilde c}_b]}_{\lambda} || \omega ; J M > ~,
\end{equation}
where $c_a^+$ is the creation operator of a real particle at state
$a$. The first factor $< a ||{\cal {\hat O}}_{\lambda} || b
>$ can be calculated for a given single particle basis independently of the nuclear model
\cite{Don79,Wal75}. By using the phonon operator $Q^{+,m}_{JM}$ in
Eq.(1), we obtain the following expression for neutrino reactions
via CC
\begin{eqnarray}
& &< QRPA || {\cal {\hat O}}_{\lambda } || ~ \omega ; JM
>  \\ \nonumber
= & & {\mathop\Sigma_{a \alpha^{'} b \beta^{'}}}  [ {\cal N}_{a
\alpha^{'} b \beta^{'} } < a \alpha^{'} || {\cal {\hat
O}}_{\lambda}  || b \beta^{'}
>  ~[ u_{pa \alpha^{'}} v_{nb
\beta^{'}} X_{a \alpha^{'} b \beta^{'}} + v_{pa \alpha^{'}} u_{nb
\beta^{'}} Y_{a \alpha^{'} b \beta^{'}} ]~,
\end{eqnarray}
where $ {\cal N}_{a \alpha^{'} b \beta^{'}} (J) =  {\sqrt{ 1 -
\delta_{ab} \delta_{\alpha^{'}  \beta^{'} } (-1)^{J + T} }}/ ({1 +
\delta_{ab}\delta_{\alpha^{'}  \beta^{'} } }) $. This form is also
easily reduced to the result by the pnQRPA which does not include
the np pairing correlations \cite{Ring08}
\begin{equation} < QRPA || {\cal {\hat O}}_{\lambda } || ~ \omega ; JM
>  = {\mathop\Sigma_{ap bn}}  [ {\cal N}_{a p b n } < a p || {\cal
{\hat O}}_{\lambda}  || b n> ~[ u_{pa} v_{nb} X_{a p b n} + v_{pa
} u_{nb } Y_{a p b n} ]~.
\end{equation}

Since NC reactions for $^{139}$La and $^{181}$Ta occur at odd-even
nuclei, we need to properly describe the ground state of odd-even
nuclei. The standard QRPA treats the ground state of the even-even
nuclei as the BCS vacuum, so that it is not easily applicable to
the reaction on these odd-even nuclei.

Here we present briefly our formalism based on the quasi-particle
shell model (QSM) \cite{suho06} to deal with such NC reactions.
First, we generate low energy spectra of odd-even nuclei by
operating one quasi-particle to the even-even nuclei constructed
by the BCS theory, {\it i.e.} $| \Psi_i > = a_{i {\mu}^{'}}^+ |
BCS
>$ and $| \Psi_f
> = a_{f {\nu}^{'}}^+ | BCS >$. Then weak transitions by NC are
calculated as
\begin{eqnarray}
& & {\mathop\Sigma_{i {\mu}^{'} f {\nu}^{'}  }}< J_f || {\cal
{\hat O}}_{\lambda } ||J_i>  \\ \nonumber & = & {\mathop\Sigma_{
{\mu}^{'} f {\nu}^{'}}} {[ \lambda ]}^{-1} {\mathop\Sigma_{ a b}}
< a p || {\cal {\hat O}}_{\lambda}
|| b p>  < J_f ||  {[ c_{ap}^+ {\tilde c}_{bp} ]}_{\lambda} ||J_i> ~  \\
\nonumber &  = & {\mathop\Sigma_{  {\mu}^{'} f {\nu}^{'}}} [ < f p
|| {\cal {\hat O}}_{\lambda} || i p> ~ u_{f p {\nu}^{'}} u_{i p
{\mu}^{'}} + {(-)}^{j_a + j_b + \lambda } < i p || {\cal {\hat
O}}_{\lambda} || f p> ~ v_{i p {\mu}^{'}} v_{f p {\nu}^{'}} ] + (p
\rightarrow n)~,
\end{eqnarray}
where we used
\begin{eqnarray}
< J_f |  {[ c_{ap}^+ {\tilde c}_{bp} ]}_{\lambda} |J_i> & = & <
BCS | a_{f {\nu}^{'}} {[ c_{ap}^+ {\tilde c}_{bp} ]}_{\lambda}
a_{i {\mu}^{'}}^+ | BCS>  \\ \nonumber &=& < J_f ||  {[ c_{ap}^+
{\tilde c}_{bp} ]}_{\lambda} ||J_i> {(-)}^{J_i + M_i } {{\hat
J}_i}^{-1} \delta_{\Lambda J_i} \delta_{M_i - \mu  } \\ \nonumber
& = & [ u_{a p {\alpha}^{'}} u_{b p {\beta}^{'}} \delta_{f
{\nu}^{'} a {\alpha}^{'}} \delta_{b {\beta}^{'} i {\mu}^{'}} +
{(-)}^{j_a + j_b + \lambda } u_{a p {\alpha}^{'}} u_{p p
{\beta}^{'}} \delta_{f {\nu}^{'} a {\alpha}^{'}} \delta_{b
{\beta}^{'} i {\mu}^{'}}]~.
\end{eqnarray}
The weak current operator is comprised by longitudinal, Coulomb,
electric and magnetic operators, ${\hat O}_{\lambda }$, detailed
at Ref. \cite{Ch09-2}. Finally, with the initial and final nuclear
states, cross sections for $\nu ({\bar \nu}) - A$ reactions
through the weak transition operator are directly calculated by
using the formulas at Ref. \cite{Wal75}. For CC reactions we
multiplied Cabbibo angle $cos ^2 \theta_c$ and took account of the
Coulomb distortion of outgoing leptons \cite{Suzuki06,Ring08}.

Our QRPA includes not only proton-proton and neutron-neutron
pairing but also neutron-proton (np) pairing correlations. But the
contribution by the np pairing is shown to be only within 1 $\sim$
2 \% for the weak interaction on $^{12}$C, such as $\beta^{\pm}$
decay and the $\nu - ^{12} $C reaction \cite{Ch09-1,Ch09-2}. Such
a small effect is easily understood because the energy gap between
neutron and proton energy spaces in such a light nucleus is too
large to be effective. But in medium-heavy nuclei, such as
$^{56}$Fe and $^{56}$Ni, the np pairing effect accounts for 20
$\sim$ 30 \% of total cross sections \cite{Ch09-2}. Therefore, in
heavy nuclei considered in this work, the np pairing could be one
of important ingredients to be considered.

Since $^{180}$Ta is a well known deformed nucleus, one needs to
explicitly consider the deformation with the Nilsson deformed
basis. But, since our QRPA is based on the spherical symmetry, we
take a phenomenological approach for the deformation in $^{180}$Ta
\cite{Ch93}.

The np pairing has two isospin contributions, T = 1 and T = 0,
which correspond to J = 0 and J = 1 pairings, respectively. Since
the J = 0 (T = 1) pairing takes the coupling of a state and its
time reversed state, radial shape is almost spherical, so that the
J = 0 (T = 1) np pairing can be included even in the spherical
symmetric model. But, most calculations by the RPA, QRPA and Shell
models considered only T = 1 pairing by neutron-neutron (nn) and
proton-proton (pp) pairings correlations. Since our framework is
based on the Hartree Fock Bogoliubob (HFB) theory, all possible T
= 1 pairings by np, nn and pp pairing correlations can be taken
into account properly.

The J = 1 (T = 0) np pairing, which is partly associated with the
tensor force, leads to the non-spherical property, {\it i.e.} the
deformation. Therefore, the J = 1 (T = 0) np coupling cannot be
included in the spherical symmetric model, in principle. However,
if we use a renormalized strength constant for the np pairing,
$g_{np}$, as a parameter to be fitted for the empirical pairing
gaps $\delta_{np}^{emp.}$, the J = 1 (T = 0) pairing can be
incorporated implicitly even in the spherical symmetric model
because the fitted $g_{np}$ may include effectively the
deformation in the nucleus.

Empirical np pairing gap $\delta_{np}^{emp.}$ is easily extracted
from the mass excess data. Theoretical pairing gap
$\delta_{np}^{th.}$ is calculated as the difference of total
energies with and without np pairing correlations \cite{Ch93}
\begin{equation}
\delta_{np}^{th.} = - [ (H_0^{'} + E_1^{'} + E_2^{'}) - (H_0 + E_1
+ E_2) ],
\end{equation}
where $H_0^{'}(H_0)$ is the Hartree-Fock energy of a ground state
with (without) np pairing and $E_1^{'} + E_2^{'} ( E_1 + E_2)$ is
a sum of the lowest two quasi-particles energies with (without) np
pairing correlations. More detailed discussion is given at Ref.
\cite{Ch93}.

But the renormalization constant $g_{np}$ may deviate largely from
$g_{np}$ = 1.0, while the deformed effects are reasonably included
by this approach. In the model including the deformation
explicitly, for example, the deformed QRPA model \cite{Saleh09},
the value may be only scattered slightly from the $g_{np}$ = 1.0.

\section{Results}
In Fig.1, we show results for NC reactions on odd-even nuclei,
$^{139}$La$( { \nu_e} , { \nu_e}^{'}) ^{139}$La$^{*}$ and
$^{181}$Ta$ ( { \nu_e} , { \nu_e}^{'}) ^{181}$Ta$^{*}$. Since both
nuclei have ${7/2}^+$ ground states and have 3/2 $\sim$ 9/2 states
for $^{139}$La and 7/2 $\sim$ 15/2 states for $^{181}$Ta as
excited states, higher multipole transitions between ground and
excited states are possible. But the GT ($1^{+}$) transition
dominates the cross sections for $^{139}$La. For $^{181}$Ta,
$2^{-}$ transition emerges to be dominant with the GT ($1^{+}$)
transition.

In the case of even-even nuclei, such as $^{12}$C and $^{56}$Ni,
NC reactions are dominated by the GT transition
\cite{Ch09-1,Ch09-2}. In specific, below 40 MeV region, neutrino
cross sections are fully ascribed to the GT transition. But with
the higher energy, other contributions such as isospin analogue
state (IAS) and spin dipole resonance (SDR) transitions account
for 30 $\sim$ 40 \% of total cross sections, as shown at figures
2,4 and 5 in Ref. \cite{Ch09-2}. Therefore, results for $^{139}$La
show a tendency similar to those of even-even nuclei. But, for
$^{181}$Ta, one could see a significant contribution by the $2^-$
transition, which becomes larger than the contribution by the GT
transition, with the higher incident energy.

Ground and excited states generated by $| \Psi_f
> = a_{f {\nu}^{'}}^+ | BCS : A (e-e)>$ reproduce experimental spectra of the lower excited
states. For both nuclei, $^{40}$Ca is used as a core. One more
point to be noticed is that we did not find any discernible
differences between $\nu_e$ and ${\bar \nu}_e$ reactions. It means
that main contributions for both reactions stem from the Coulomb
and longitudinal parts because the difference originates from the
magnetic and electric interference term \cite{Wal75}.


Fig.2 shows results for CC reactions on even-even nuclei,
$^{138}$Ba$ ( \nu_e , e^{-}) ^{138}$La$^{*}$ and $^{180}$Hf$ (
\nu_e , e^{-}) ^{180}$Ta$^{*}$, respectively. Main contribution
for CC reactions on both nuclei is the GT transition below 40 MeV
region, and remained contributions, 20 $\sim$ 30 \% of total cross
sections, are ascribed to other transitions, such as IAS and SDR
($1^-, 0^{\pm}, 3^{\pm}$ and $2^+$). With the increase of incident
energy, those contributions are increased to 60 $\sim$ 70 \%,
which are almost 2 times of those in NC reactions. These roles of
the GT and other transitions are also typical of CC neutrino
reactions on even-even nuclei, for example, $^{12}$C and $^{56}$Fe
as shown in figure 1 and 6 at Ref. \cite{Ch09-2}.

The Fermi function for the Coulomb distortion is used on the whole
energy region in Fig.2. Strength constants of the pairing
correlations, $g_{nn} = 1.1667(1.441)$, $g_{pp} = 0.950(0.899)$
and $g_{np} = 2.3435(2.9915)$, are adjusted to the empirical
pairing gaps $\Delta_{nn} =0.883(0.712)$ MeV, $\Delta_{pp} =
1.087(1.065)$ MeV and $\delta_{np} =0.262(0.192)$ MeV for
$^{138}$Ba ($^{180}$Hf), respectively \cite{Ch93}. These strength
parameters should be understood as renormalized constants
introduced to consider finite Hilbert particle model spaces used
here. Therefore, they deviated a little bit from 1.0. But, the
$g_{np}$ value for $^{180}$Hf is quite larger than the $g_{np}$
for $^{138}$Ba because of the deformation as explained above.

In experimental sides, we do not have any data for neutrino
reactions on these nuclei. But a recent experiment \cite{RCNP} for
the GT transition by the ($^{3}$He, t) reaction could help us to
constrain theoretical estimations of neutrino reactions via CC.
Moreover, Ref. \cite{RCNP} deduced neutrino reaction data from the
measured GT strength distribution by using the recipe on Ref.
\cite{Qian97} and compared to the theoretical calculations by the
RPA.

In the left panels of Fig.3, we present our results for the GT(--)
strength distribution for $^{138}$La and $^{180}$Ta, which are
calculated as
\begin{equation}
B(GT_{\pm}) = { 1 \over  {2 J_i + 1  }} {| < f ||
{\mathop\Sigma_{k}  \sigma_k \tau_{k_{\pm}}} || i >|}^2 ~.
\end{equation}
Various peak positions in lower energy states are confirmed to be
well suited to the experimental data \cite{RCNP} performed below
10 MeV region. Contributions above nucleon thresholds, in
particular, around 20 MeV region, are found to play significant
roles in relevant neutrino reactions.

Running sums of the GT(--) strength distribution up to 10 MeV are
shown in the right panels. Our results for the running sum up to 8
MeV are 5.5 and 3.8 for $^{138}$La and $^{180}$Ta, if we take the
universal quenching factor $f_q= 0.74$ for the axial coupling
constant $g_A$ into account. They reproduce well the experimental
data, $5.8 \pm 1.6$ and $4.4 \pm 0.9$, respectively, reported at
Ref. \cite{RCNP}. For other multipole transitions, we did not use
the quenching factor by following the discussions at Ref.
\cite{Suzuki09}. More data on the higher energy region for the GT
and other spin-isospin excitations by the $(^{3}$He,t) or (p,n)
reactions could give more reliable information on the nuclear
structure for the relevant weak transitions.

In table 1, our full calculations for CC cross sections weighted
by the assumed neutrino spectra are tabulated with the
semi-empirical data deduced from the $(^{3}$He,t) reaction and the
RPA calculations. To compare with the experimental data, we
exploited the Fermi function below 40 MeV and the effective
momentum approach (EMA) above 40 MeV for the Coulomb correction
\cite{Ring08,Ch09-1}. Results for ${^{138}}$La are more or less
consistent with the semi-experimental data, if we consider the
inherent error bars in the data, which arise from the ambiguity on
the GT strength obtained from the experiment and the uncertainty
in the theoretical deduction of the corresponding cross sections
from the data.
\begin{table}
\caption[bb]{CC cross sections in a unit of 10$^{-42} cm^2$ for
$^{138}$Ba$ ( \nu_e , e^{-}) ^{138}$La$^{*}$ and $^{180}$Hf$ (
\nu_e , e^{-}) ^{180}$Ta$^{*}$ reactions. They are averaged by the
neutrino flux in a core collapsing SN for a given temperature.
Experimental data and RPA results are cited from Ref.\cite{RCNP}.}
\vskip0.5cm \setlength{\tabcolsep}{3.0 mm}
\begin{tabular}{ccccccc}\hline
                       T(MeV)     & $^{138}$La  &
           &  & $^{180}$Ta & &       \\
                                 \hline\hline
     & Exp.  & RPA & Ours & Exp. & RPA & Ours      \\
   4 & 74 & 61 & 68 & 151 & 115 & 76
     \\ \hline
  6  & 226 & 156 & 254 & 399 & 272 & 316 \\ \hline
  8  & 435 & 281 &  554 & 752 & 485 & 672\\ \hline\hline
\end{tabular}
\label{tab:result1}
\end{table}

Results for ${^{180}}$Ta seem to underestimate the empirical data
about 10 $\sim$ 20 \% for T = 6 $\sim$ 8 MeV. But it leads to
about 2 times difference at T = 4 MeV. Cross sections at low
temperature need to be further studied. Since our calculations are
carried out in the spherical basis, the T = 0 (J = 1) np pairing
is included implicitly by increasing the strength parameter
$g_{np}$ at the T = 1 (J = 0) np pairing contribution. But, as
well known, single particle energy states may be changed by the
deformation, which could affect relevant transitions at low
temperature. Nevertheless, our results are more advanced to the
semi-empirical data compared with previous theoretical
calculations \cite{RCNP}.

For the supernovae application, in Fig. 4, we show the cross
sections averaged by the presumed neutrino flux in a core
collapsing SN. The heavier nuclei we go to, the larger cross
sections are obtained. For the light and medium-heavy nuclei,
magnitudes of CC reactions are about 5 times larger than those by
NC. But, on the heavier nuclei, $^{138}$La and $^{180}$Ta, CC
reactions are larger about 10 times than those by NC reactions.
Applications to the $\nu$-process for the relevant nuclei are in
progress.

\section{Summaries and Conclusion}
We calculated neutrino reactions via neutral and charged currents
related to the two heaviest odd-odd nuclei, $^{138}$La and
$^{180}$Ta, by including multipole transitions up to $J^{\pi} =
4^{\pm}$ with explicit momentum dependence. Our QRPA includes
neutron-proton (np) pairing as well as neutron-neutron and
proton-proton pairing correlations. Since energy gaps between
proton and neutron energy spaces in heavy nuclei are adjacent to
each other, the np pairing may affect significantly the nuclear
weak interaction. We included explicitly the T = 1 (J = 0) np
pairing. The T = 0 (J = 1) contribution in the np pairing, which
is believed to cause the deformation of $^{180}$Ta, is included
implicitly by incresing the strength parameter $g_{np}$ at the T =
1 (J = 0 ) pairing matrix element to reproduce the empirical np
pairing gap.

To describe the NC reactions on odd-even nuclei, $^{139}$La and
$^{181}$Ta, we exploited the quasi-particle shell model (QSM).
Their results are consistent with the trend of NC reactions on
even-even nuclei, which are fully dominated by the GT transition
below 40 MeV with the increase of other transitions above 40 MeV
region. But, for $^{181}$Ta, the GT dominance is relatively
weakened compared to results for other even-even nuclei because of
the deformation peculiar to the nucleus.

In the CC reactions on even-even nuclei, such as $^{12}$C,
$^{56}$Fe and $^{56}$Ni, about 60 \% of cross sections is ascribed
to the GT transition in the energy region below 40 MeV.
Contributions by the IAS and SDR transitions become much larger
than those in NC reactions. Namely, the GT transition in CC
reactions is not so dominant as that of NC reactions. Our results
for CC reactions ,$^{138}$Ba$ ( \nu_e , e^{-}) ^{138}$La and
$^{180}$Hf$ ( \nu_e , e^{-}) ^{180}$Ta, also show such a tendency
typical of the CC reactions on even-even nuclei.

Recent experimental data carried out at RCNP \cite{RCNP}, the
Gamow-Teller (GT) strength distributions and their total strength,
are well reproduced with the universal quenching factor and a
proper choice of the Coulomb correction. Neutrino cross sections
averaged by the neutrino flux emitted from core collapsing
supernovae are also compared with the semi-empirical results
deduced from the GT strength data. Our results are found to be
more consistent with the semi-empirical data than previous RPA
calculations. But, for $^{180}$Ta, which is a well known deformed
nucleus, more careful refinement is necessary. Since the exotic
nuclei of astrophysical importance are subtly deformed, we need to
develop the Deformed QRPA (DQRPA) which explicitly includes the
deformation in the Nilsson basis under the axial symmetry
\cite{Ha10}.

Since our theoretical data for relevant neutrino reactions are
shown to be consistent with the recent empirical data, nuclear
abundances of $^{138}$La and $^{180}$Ta would not be changed so
much enough to affect previous predictions. Abundance of
$^{138}$La is more or less reproduced by considering CC reactions
$T_{\nu_e}$ = 4 MeV neutrino as claimed previously, but $^{180}$Ta
is still overproduced. Linkage transitions between ground and
isomer states in $^{180}$Ta may be one of the reasonable solutions
to explain the overproduction properly \cite{Haya10}.

The QRPA is a very efficient method to consider multi-particle and
multi-hole interactions and their configuration mixing, and
successfully described nuclear reactions sensitive on the nuclear
structure, such as $2 \nu 2 \beta$ and 0$\nu 2 \beta$ decays.
Possible ambiguities on the neutrino-induced reaction caused by
the nuclear weak structure can be reduced by reproducing the
available data related to the nuclear $\beta$ decay as well as the
forthcoming GT transition data by ($^{3}$He, t) or (p,n) reactions
\cite{Shim10}.

\section*{ACKNOWLEDGMENTS}
This work was supported by the National Research Foundation of
Korea (2009-0077273) and one of author, Cheoun, was supported by
the Soongsil University Research Fund. This work was also
supported in part by Grants-in-Aids for Scientific Research
(20244035, 20105004) of Japan.

\newpage

\newpage

\begin{figure}
\includegraphics[width=0.85\linewidth]{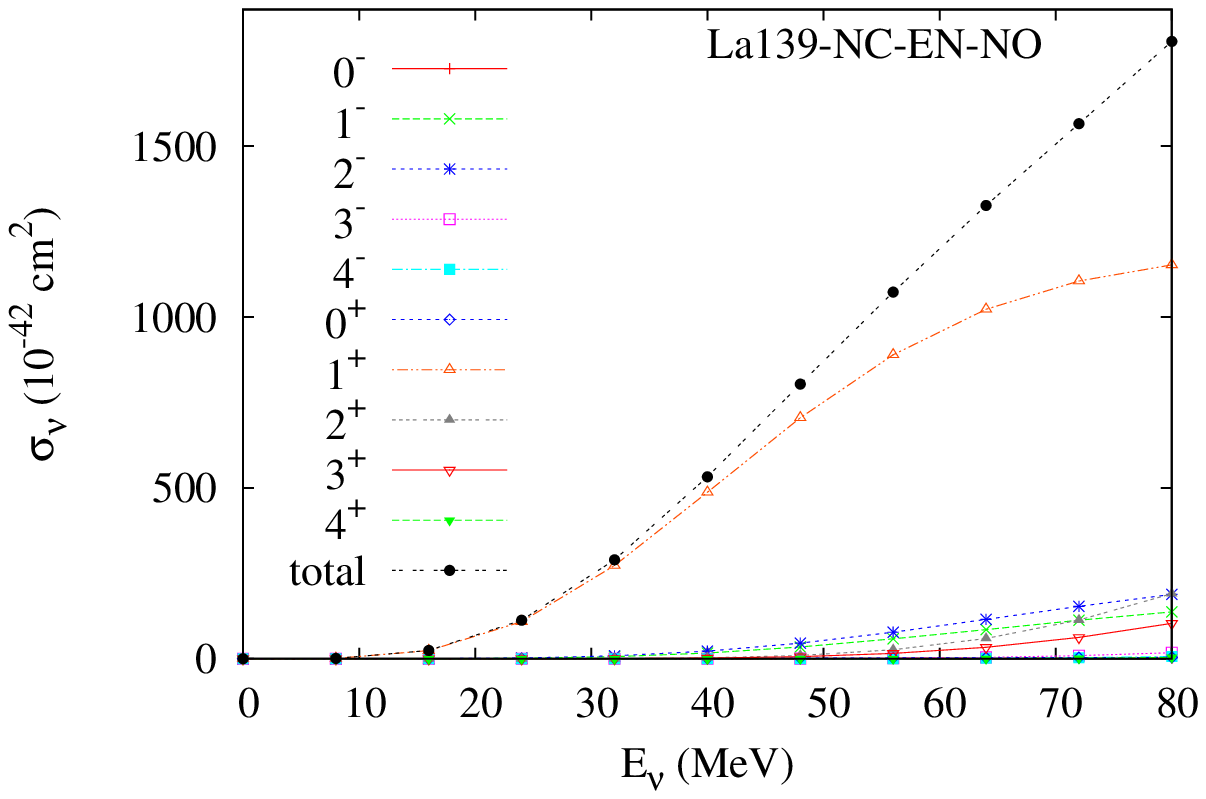}
\includegraphics[width=0.85\linewidth]{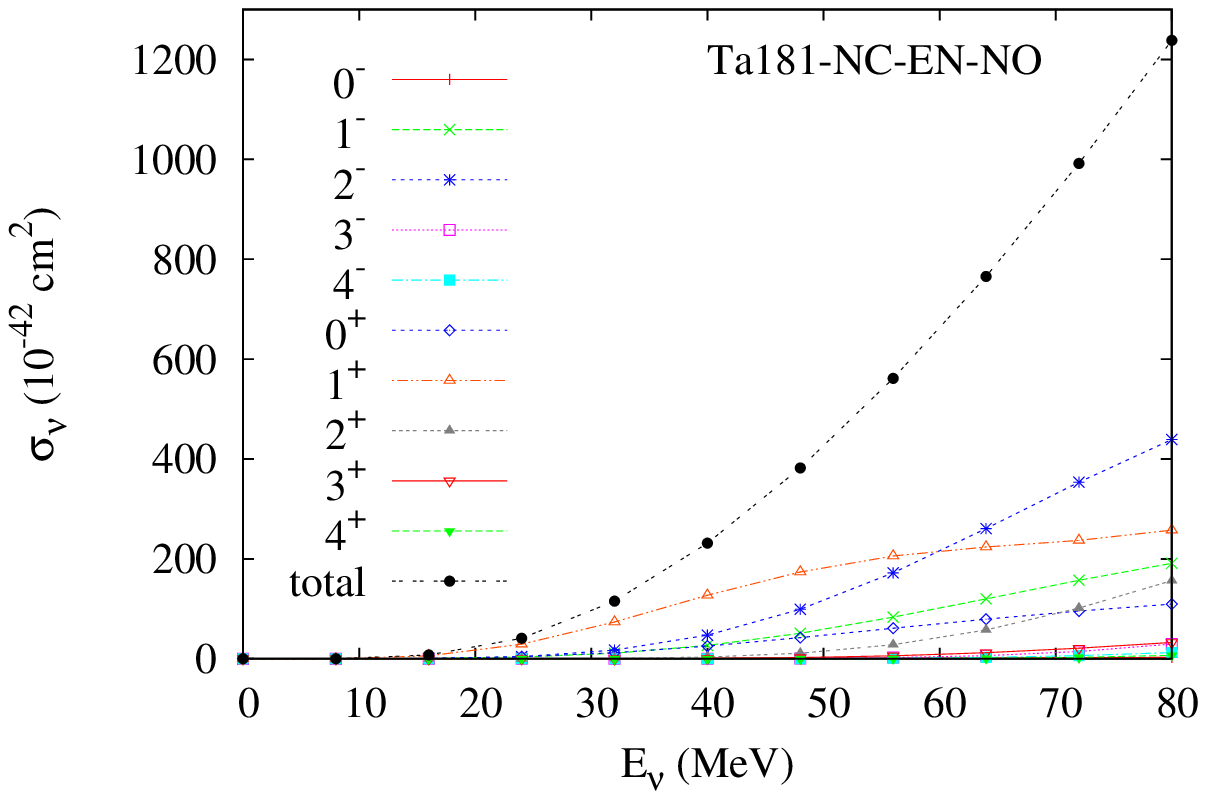}
\caption{(Color online) Cross sections by neutral current,
$^{139}$La$ ( { \nu}_e , { \nu}^{'}_e) ^{139}$La$^*$ and
$^{181}$Ta$ ( { \nu_e} , {\nu_e}^{'}) ^{181}$Ta$^{*}$ reactions
for $J_{\pi} = 0^{\pm} \sim 4^{\pm}$ states. Transition matrix
elements are calculated by the QSM, Eq.(6).} \label{fig1}
\end{figure}

\begin{figure}
\includegraphics[width=0.85\linewidth]{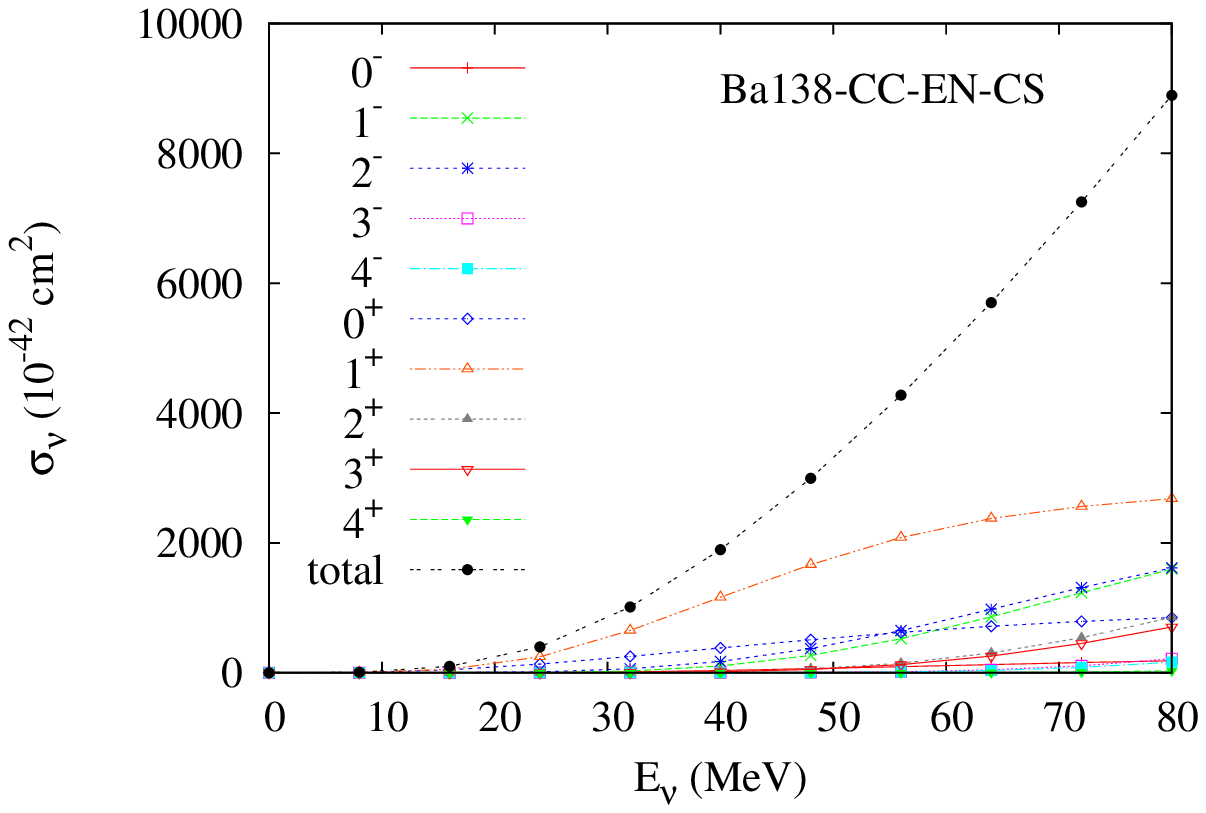}
\includegraphics[width=0.85\linewidth]{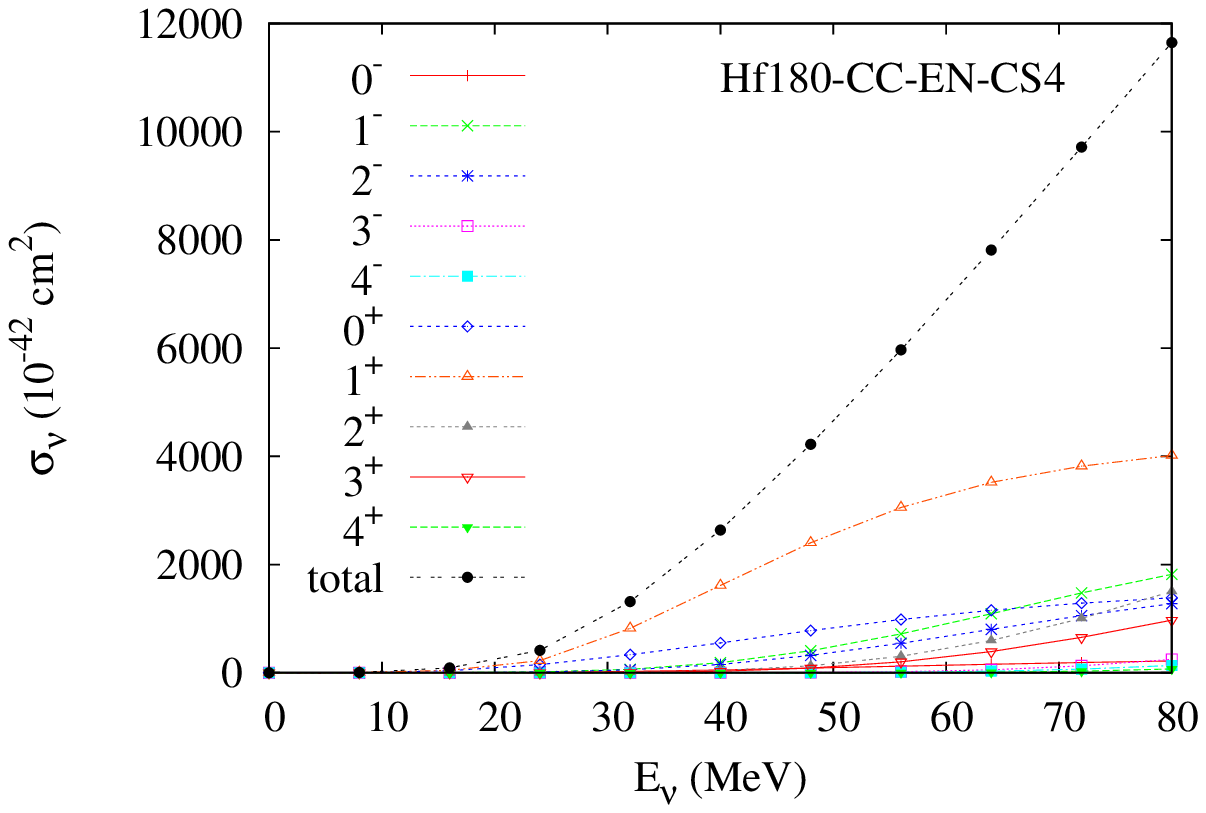}
\caption{(Color online) Cross sections by charged current,
$^{138}$Ba$ ( \nu_e , e^{-}) ^{138}$La$^{*}$ and $^{180}$Hf$ (
\nu_e , e^{-}) ^{180}$Ta$^{*}$, for $J_{\pi} = 0^{\pm} \sim
4^{\pm}$ states. Transition matrix elements are calculated by the
QRPA, Eq.(4).} \label{fig2}
\end{figure}

\begin{figure}
\centering
\includegraphics[width=7.8cm]{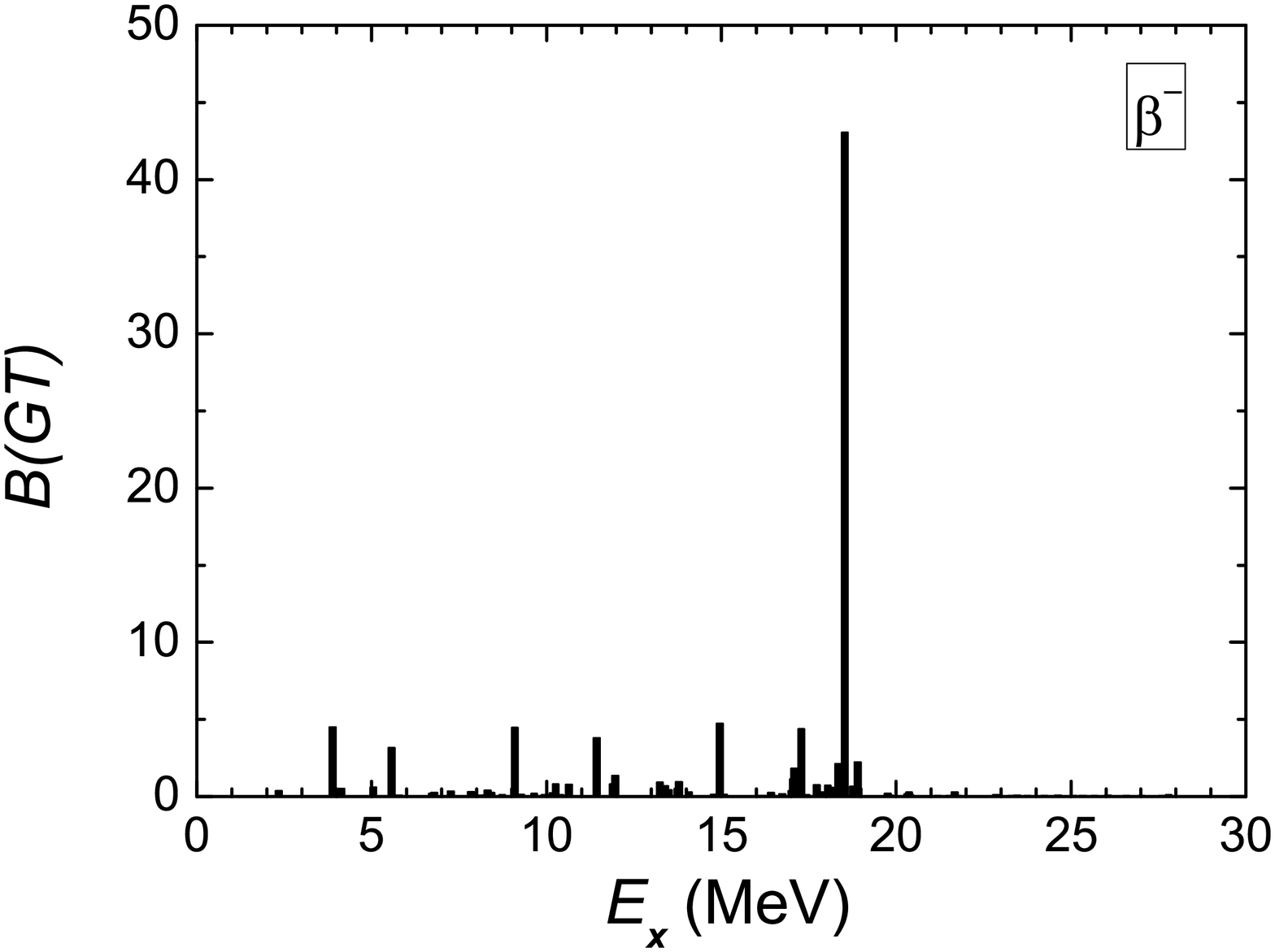}
\includegraphics[width=7.8cm]{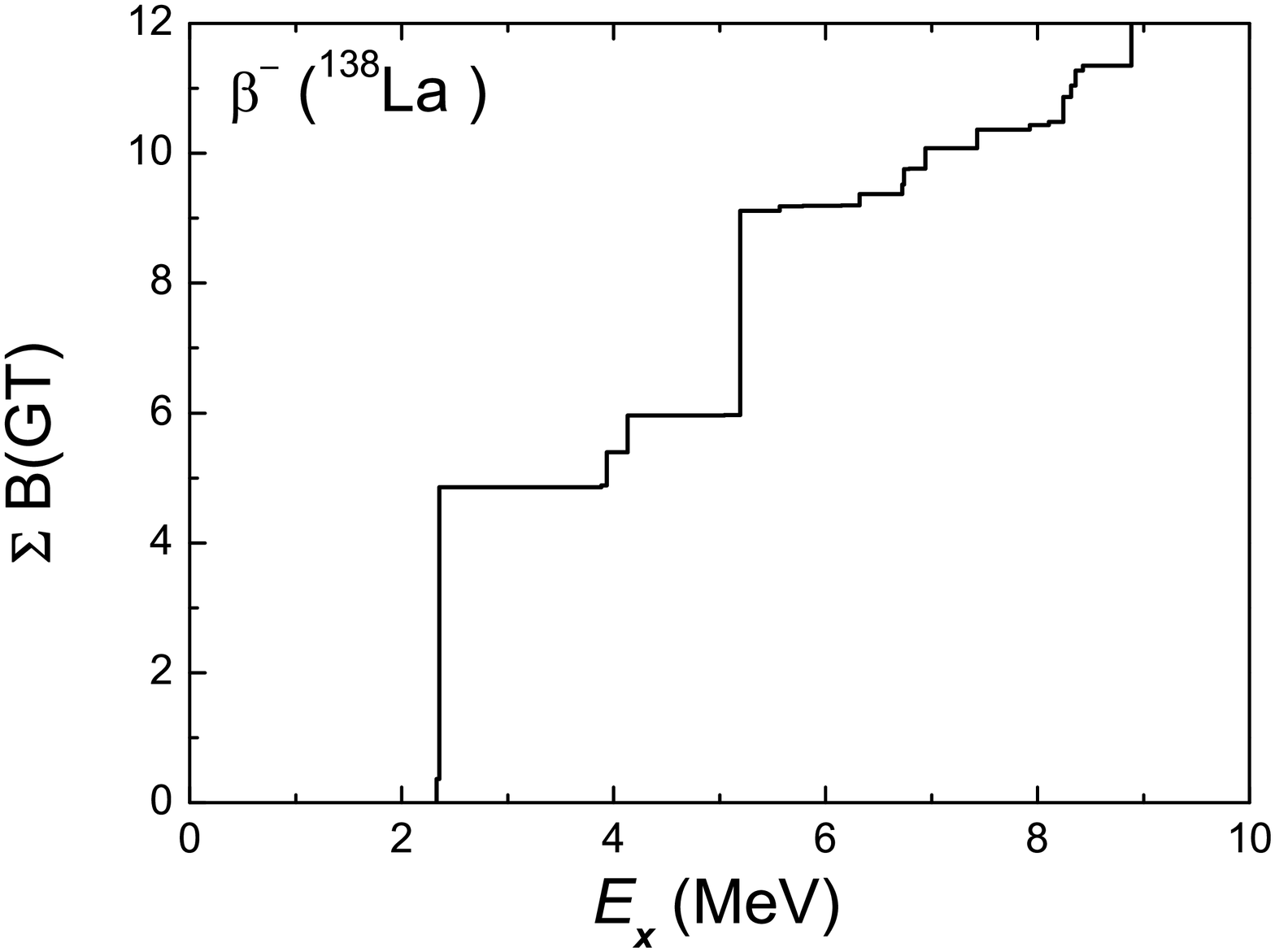}
\includegraphics[width=7.8cm]{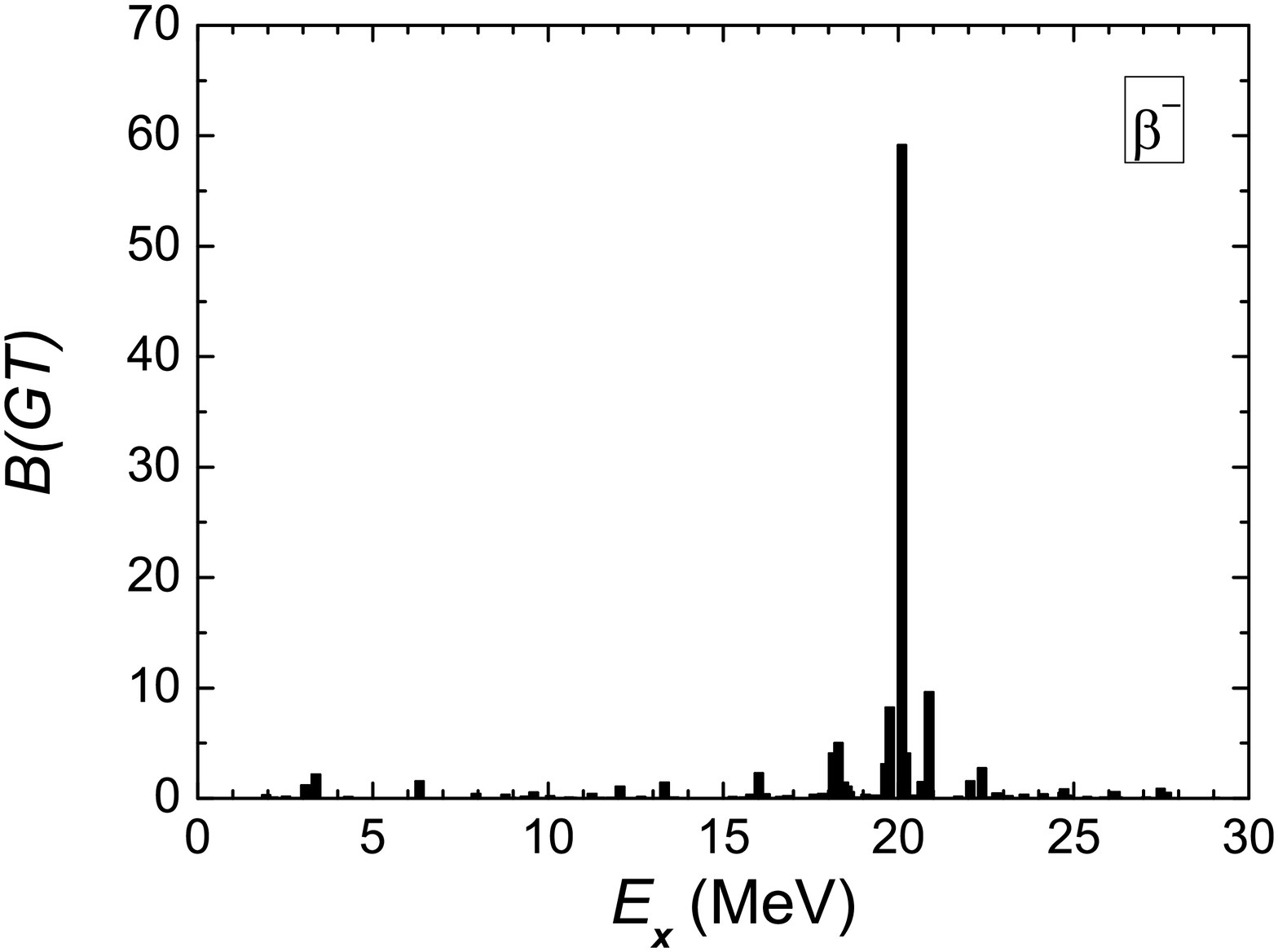}
\includegraphics[width=7.8cm]{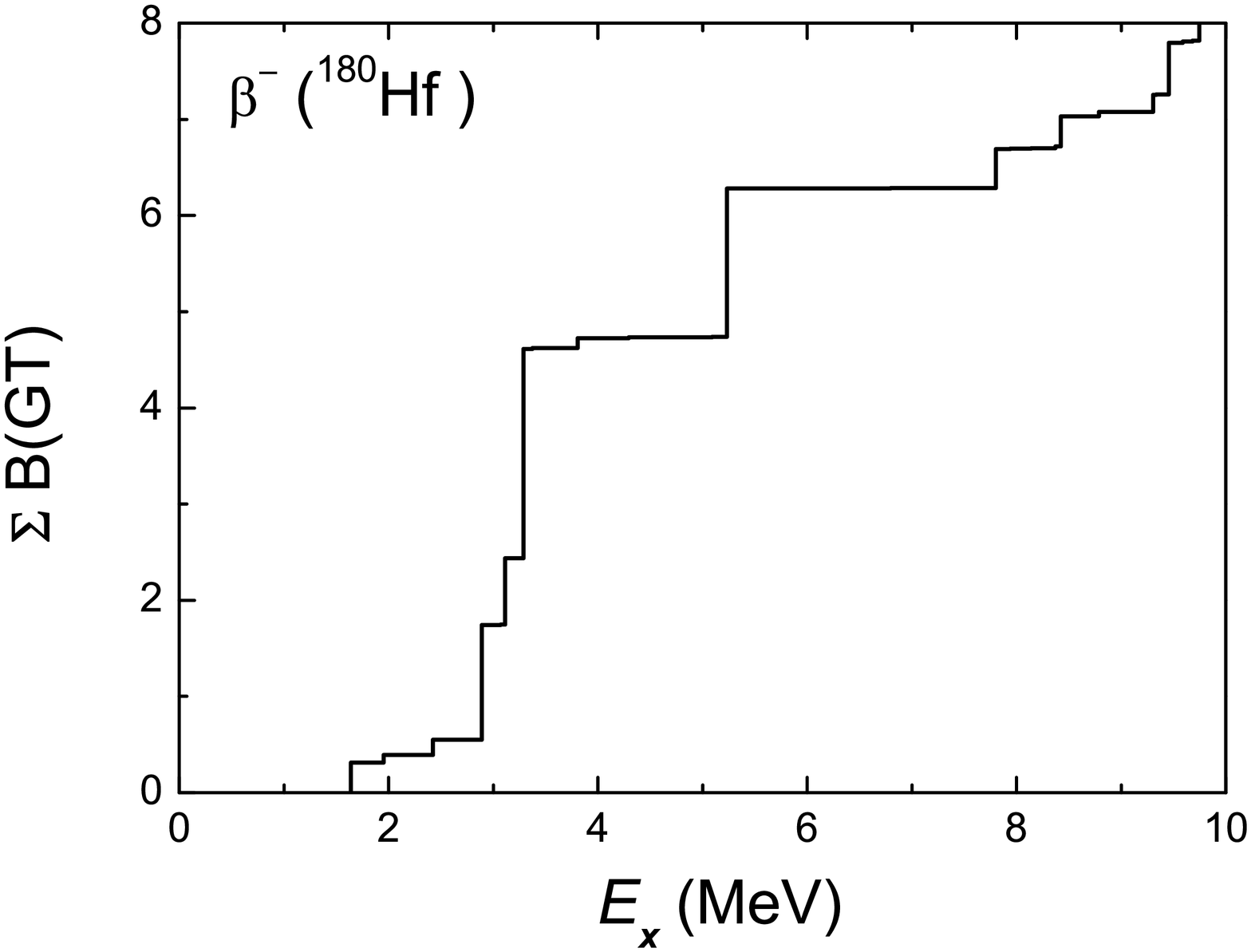}
\caption{ The Gamow-Teller strength distribution $B(GT_{-})$ by
Eq.(9) and their running sums for $^{138}$Ba$\rightarrow ^{138}$La
(upper panels) and $^{180}$Hf$\rightarrow ^{180}$Ta (lower
panels). The universal quenching factor $f_q$ = 0.74 is not used
for the results in these figures.}\label{fig3}
\end{figure}

\begin{figure}
\includegraphics[width=0.85\linewidth]{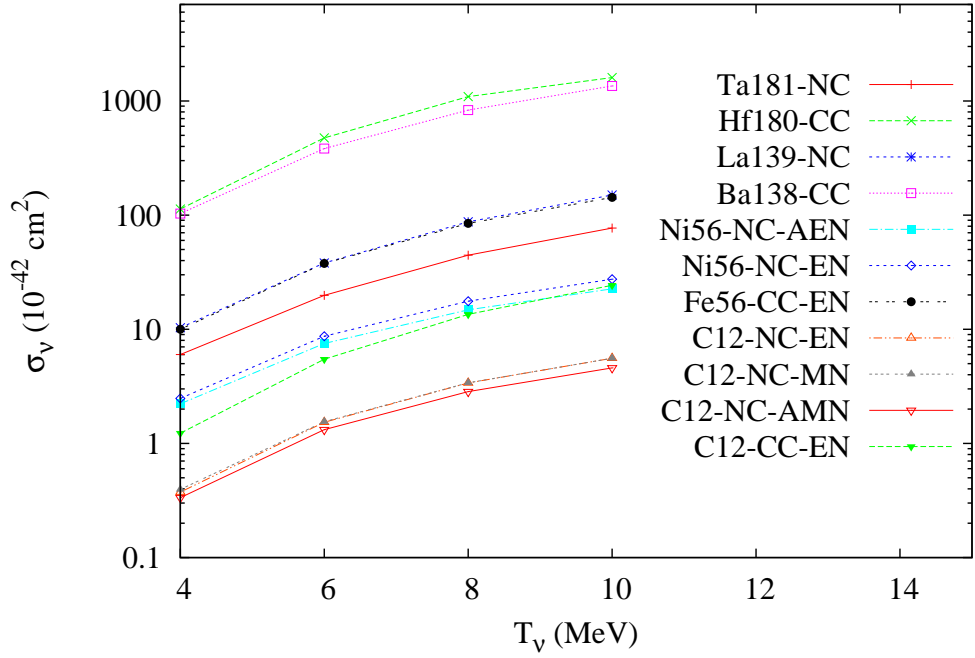}
\caption{(Color online) Temperature dependence of the energy
weighted cross section for $\nu - A$ reactions, whose neutrino
spectra for the supernovae are assumed as the Fermi-Dirac
distribution. Results for $^{12}$C, $^{56}$Fe and $^{56}$Ni are
referred from our previous calculations \cite{Ch09-1,Ch09-2}}
\label{fig4}
\end{figure}


\begin{thebibliography}{140}
\par
\vskip0.5cm
\par
\def\pr{Phys. Rev.}
\def\prl{Phys. Rev. Lett.}
\def\nc{Nucl. Phys.}
\def\pl{Phys. Lett.}
\def\nuc{Nuovo. Cim.}
\def\pro{Prog. Theo. Phys}
\def\so{Sov.J.Nucl.Phys.}
\def\can{Can. J. Phys.}
\par

\bibitem{Beer81} H. Beer and R. A. Ward, Nature {\bf 291}, 308 (1981).
\bibitem{Yokoi83} K. Yokoi and K. Takahashi, Nature {\bf 305}, 198 (1983).
\bibitem{Woosley78} S. E. Woosley,  W.M. Howard, Astrophys. J. Suppl. {\bf 36}, 285 (1978).
\bibitem{Woosley90} S. E. Woosley, D. H. Hartmann, R. D. Hoffmann,
and W. C. Haxton, Astrophys. J. {\bf 356}, 272 (1990).

\bibitem{yoshida08} T. Yoshida, T. Suzuki, S, Chiba, T. Kajino, H.
Yokomukura, K. Kimura, A. Takamura, H. Hartmann, Astro. Phys. J.
{\bf 686}, 448 (2008).
\bibitem{Kolbe03-a} E. Kolbe, K. Langanke, G Martinez-Pinedo and P. Vogel, J. Phys.
G {\bf 29}, 2569 (2003).






\bibitem{Suzuki06} T. Suzuki, S. Chiba, T. Yoshida, T. Kajino, T.
Otsuka, Phys. Rev. C {\bf 74}, 034307 (2006).
\bibitem{Ring08} N. Paar, D. Vretenar, T. Marketin, and P. Ring,
Phys. Rev. {\bf C 77}, 024608 (2008).

\bibitem{Heg} A. Heger, E. Kolbe, W. C. Haxton, K. Langanke, G. Mart{\'i}nez-Pinedo,
S. E. Woosley, Phys. Lett. {\bf B606}, 258 (2005).


\bibitem{Suzuki09} T. Suzuki, M. Honma, K. Higashiyama, T. Yoshida, T. Kajino, T.
Otsuka, H. Umeda, and K. Nomoto, Phys. Rev. C {\bf 79}, 061603(R)
(2009).


\bibitem{Wana06} Shinya Wanajo, Astrophys. J. {\bf 647}, 1323 (1990).



\bibitem{RCNP} A. Byelikov {\it et. al.}, Phys. Rev. Lett. {\bf
98}, 082501 (2007).
%
\bibitem{Haya} T. Hayakawa, T. Shizuma, T. Kajino, K. Ogawa, and H. Nakada,
Phys. Rev. {\bf C 77}, 065802 (2008).
\bibitem{Anders89} E. Anders and N. Grevesse, Geochim. Cosmochim.
Acta {\bf 53}, 197 (1989).

\bibitem{Haya10} T. Hayakawa, T. Kajino, S. Chiba, and G. J. Mathews,
Phys. Rev. {\bf C 81}, 052801(R) (2010).

\bibitem{Ch09-2} Myung-Ki Cheoun, Eunja Ha, K. S. Kim and T.
Kajino, J. Phys. {\bf G 37}, 055101, (2010).
\bibitem{Ch09-1} Myung-Ki Cheoun, Eunja Ha, S. Y. Lee, W. So, K. S. Kim and T.
Kajino, Phys. Rev. {\bf C 81}, 028501 (2010).

\bibitem{Ch93} M. K. Cheoun, A. Bobyk, Amand Faessler, F. Simcovic and
G. Teneva, {\nc} {\bf {A561}}, 74 (1993) ; {\nc} {\bf {A564}}, 329
(1993); M. K. Cheoun, G. Teneva and Amand Faessler, Prog. Part.
Nuc. Phys. {\bf 32}, 315 (1994) ; M. K. Cheoun, G. Teneva and
Amand Faessler, {\nc} {\bf A587}, 301 (1995).

\bibitem{Don79} T. W. Donnelly and W. C. Haxton, ATOMIC DATA AND
NUCLEAR DATA {\bf 23}, 103 (1979).

\bibitem{Wal75} J. D. Walecka, {\it Muon Physics}, edited by V.
H. Huges and C. S. Wu (Academic, New York, 1975), Vol II.











\bibitem{suho06} Jouni Suhonen, {\it From Nucleons and to Nucleus},
TMP, Springer-Verlag, Heidelberg, (2007).

\bibitem{Belic} D. Belic {\it et. al.}, Phys. Rev. Lett. {\bf
83}, 5242 (1999).


\bibitem{Kolbe95}E. Kolbe, K. Langanke, F.-K.Thielemann, and P.
Vogel, Phys. Rev. C {\bf 52}, 3437 (1995).



\bibitem{Kolbe03}E. Kolbe, Nucl. Phys. {\bf A719}, 135c (2003).



\bibitem{Ryck02} N. Jachowicz, K. Heyde, J.
Ryckebusch, and S. Rombouts, Phys. Rev. C {\bf 65}, 025501 (2002).




\bibitem{Saleh09} Mohamed Saleh Yousef, Vadim Rodin, Amand Faessler, Fedor Simkovic,
Phys. Rev. {\bf C 79}, 014314 (2009).

\bibitem{Qian97} Y. Z. Qian, W. C. Haxton, K. Langanke, P. Vogel, Phys. Rev. {\bf C 55},
1532 (1997).


\bibitem{Ha10} Eunja Ha and Myung-Ki Cheoun, {\it The 10th International Symposium on
Origin of Matter and Evolution of Galaxies}, edited by I. Tanihara
{\it et al.}, AIP, New York, 351 (2010).

\bibitem{Shim10} Y. Shimbara {\it et al.}, {\it The 10th International Symposium on
Origin of Matter and Evolution of Galaxies}, edited by I. Tanihara
{\it et. al.}, AIP, New York, 201 (2010).
\end{thebibliography}
\end{document}